\documentclass{aa}
\usepackage{graphicx}
\usepackage{natbib}

\begin{document}
   \title{Hard X-ray emission from a young massive star-forming cluster}
   \titlerunning{X-ray emission in massive star-forming regions}
   \authorrunning{Beuther et al.}
%   \subtitle{I. Overviewing the $\kappa$-mechanism}

   \author{H. Beuther\inst{1}, J. Kerp\inst{2}, T. Preibisch\inst{1}, T. Stanke\inst{1}, P. Schilke\inst{1}}
   \offprints{H. Beuther}
   \institute{Max-Planck-Institut f\"ur Radioastronomie, Auf dem H\"ugel 69, 53121 Bonn, Germany\\ \email{beuther@mpifr-bonn.mpg.de}
         \and
          Radioastronomisches Institut der Universit\"at Bonn, Auf dem H\"ugel 71, 53121 Bonn, Germany
\thanks{Based on observations with the {\it Chandra} X-ray observatory, the Calar Alto 3.5\,m telescope and the IRAM Plateau de Bure Interferometer. {\it Chandra} is operated by the NASA and the CXC, Calar Alto by the Max-Planck-Institut for Astronomy in Heidelberg, jointly with the Spanish National Commission for Astronomy, and IRAM is supported by INSU/CNRS (France, MPG (Germany), and IGN (Spain).}
             } \date{Received: June 18, 2002; accepted: August 30, 2002}

   \abstract{We report the detection of hard X-ray emission ($>2$~keV)
   from a number of point sources associated with the very young massive
   star-forming region IRAS~19410+2336. The X-ray emission is
   detected from several sources located around the central and most
   deeply embedded mm continuum source, which 
   remains undetected in the X-ray regime. All X-ray sources have K-band
   counterparts, and those likely belonging to the evolving massive
   cluster show near-infrared colors in the 2MASS data indicative
   of pre-main-sequence stages. The X-ray luminosities around
   $10^{31}$~erg\,s$^{-1}$ are at the upper end of luminosities known for
   low-mass pre-main-sequence sources, and mass estimates based on the
   infrared data indicate that at least some of the X-ray detected
   sources are intermediate-mass objects. Therefore, we conclude that
   the X-ray emission is due to intermediate-mass pre-main-sequence
   Herbig Ae/Be stars or their precursors. The emission process is
   possibly due to magnetic star-disk interaction as proposed for their
   low-mass counterparts.  }

   \maketitle
%
%________________________________________________________________

\section{Introduction}
\label{introxray}

In the past, star formation research in the X-ray regime has focused
strongly on low-mass objects, e.g., T Tauri stars. These objects
emit mainly in the soft range of the X-ray spectrum
($<2$\,keV) with typical X-ray luminosities between
$10^{28}-10^{30}$~erg\,s$^{-1}$. The satellite observatory ROSAT was an ideal
instrument to study such stars within a few 100~pc distance
from the Sun, and the observed emission can be explained by enhanced
solar-type magnetic activity \citep{feigelson 1999}. In the last
few years, a rising number of Class I protostars, which are still
deeply embedded within their natal molecular cores ($A_{\rm{V}}
\approx 10-100$~mag), have been detected in the hard X-ray regime
between 2 and 10 keV with X-ray luminosities higher than
$10^{30}$~erg\,s$^{-1}$. These detections were mostly made with the
X-ray satellites ASCA and, most recently, with {\it Chandra}. Magnetic
star--disk interactions are thought to be the most likely explanation
for the hard X-ray emission \citep{hayashi 1996,montmerle
2000}. Furthermore, X-ray variability is observed in all types of
low-mass pre-main-sequence objects. For an excellent recent review on
these topics see
\citet{feigelson 1999}. Recently, \citet{tsuboi 2001} and
\citet{tsujimoto 2002} reported the first tentative detections of
deeply embedded Class 0 protostellar candidates in OMC3 by {\it Chandra}.

In comparison with the low-mass regime, X-ray observations of massive
star-forming regions have been rare. Due to the high visual extinction
within such regions ($A_{\rm{v}}$ up to a few 100 or even 1000), soft
X-ray emission is completely absorbed by the gas along the line of
sight. \citet{hofner 1997} detected with ASCA for the first time hard
X-ray emission in the massive star-forming region W3. Because of the
low angular resolution of ASCA ($>1'$) they could not determine
whether the emission is caused by the superposition of many point
sources, e.g., protostellar clusters, or whether it is due to a
wind-shocked cavity resulting from strong stellar winds interacting
with the surrounding medium.  Recently,
\citet{churchwell 2001} reported that {\it Chandra} data of the same
region with a spatial resolution of $0.5''$ resolve the emission into
many individual sources distributed over the entire W3 complex. Similarly,
\citet{garmire 2000} and \citet{feigelson 2002} reported about 1000 
X-ray emitting pre-main-sequence stars between 0.05 and 50 M$_{\odot}$
in the Orion Nebula. Additionally, \citet{zinnecker 1994} found X-ray
emission with the ROSAT satellite in the soft X-ray band associated
with several intermediate-mass Herbig Ae/Be pre-main-sequence
stars. The derived X-ray luminosities for the Herbig Ae/Be stars are
ranging between $10^{30}$~erg\,s$^{-1}$ and
$10^{32}$~erg\,s$^{-1}$. \citet{preibisch 1995} speculated that the
emission might originate from coronal activity due to shear dynamo
action. A study of the more distant molecular clouds Monoceros
and Rosette also found indirect evidence for X-ray emission from
intermediate-mass pre-main-sequence sources \citep{gregorio 1998}. An
example of hard X-ray emission from a Herbig Be star is MWC 297
\citep{hamaguchi 2000}. In a recent X-ray study of the Monoceros R2
molecular cloud, \citet{kohno 2002} detected X-ray emission for stars
of all masses, in particular hard X-ray emission from high-mass
pre-main-sequence or Zero-age-main-sequence stars. Based on these
results, X-ray emission seems likely to be an ubiquitous phenomenon in
the protostellar evolution of stars of all masses.

So far, X-ray studies in high-mass star-forming regions focused on
more evolved massive star formation sites, while the very deeply
embedded phase --~and thus the youngest stage of stellar evolution~--
has not been detected at all. Results obtained in recent years by our
group \citep{sridha, beuther 2002a, beuther 2002b, beuther 2002d} and
other groups studying massive star formation (e.g.,
\citealt{cesaroni 1997}, \citealt{zhang 2002},
\citealt{tan 2002}, \citealt{yorke 2002}) support the
hypothesis that massive stars form via disk accretion in a similar
fashion as low-mass stars. Therefore, high-mass star-forming cores are
promising candidate regions where in rather small spatial areas a
number of sources could be hard X-ray emitters via the physical
process of star-disk interactions \citep{hayashi 1996,montmerle
2000}. Necessary observational requirements are first of all
sensitivity in the hard X-ray regime ($>2$~keV), because only hard
X-ray photons can penetrate high gas column densities. Additionally,
high angular resolution is needed to resolve different sub-sources of
the forming cluster. While ROSAT was not sensitive to hard X-ray
photons, the spatial resolution of ASCA was not sufficient to study
massive star-forming regions in detail at their typical distances of a
few kpc.  The new-generation X-ray satellite telescopes {\it Chandra} and
XMM-Newton comprise both features, being sensitive up to 10~keV, and
having a spatial resolution of $0.5''$ ({\it Chandra}) and $15''$
(XMM-Newton), respectively. Especially {\it Chandra} is able to resolve many
different sub-sources as impressively demonstrated in Orion by
\citet{garmire 2000} and \citet{feigelson 2002}.

\begin{figure}[ht]
\includegraphics[angle=-90,width=8.7cm]{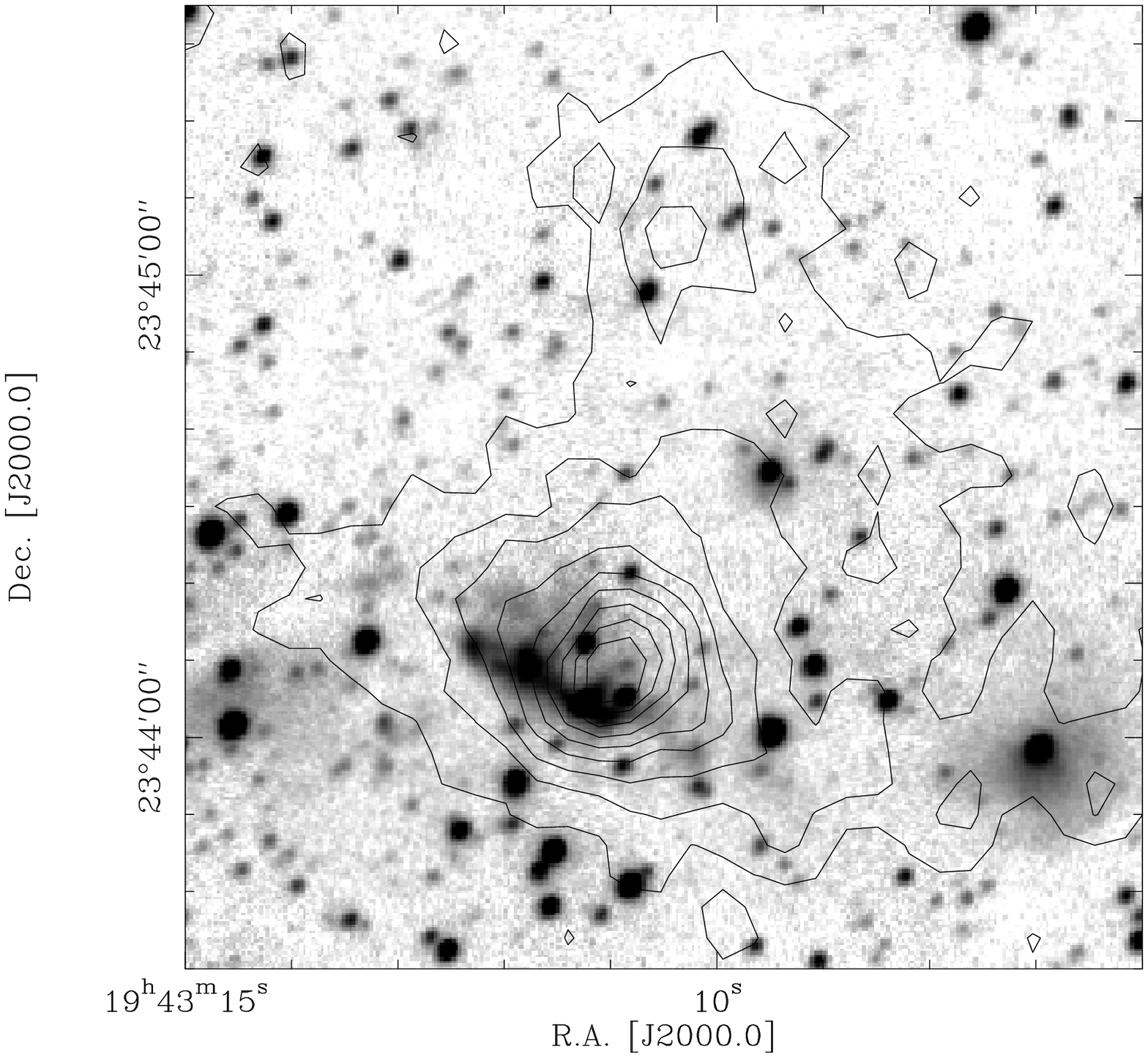}
\includegraphics[angle=-90,width=8.7cm]{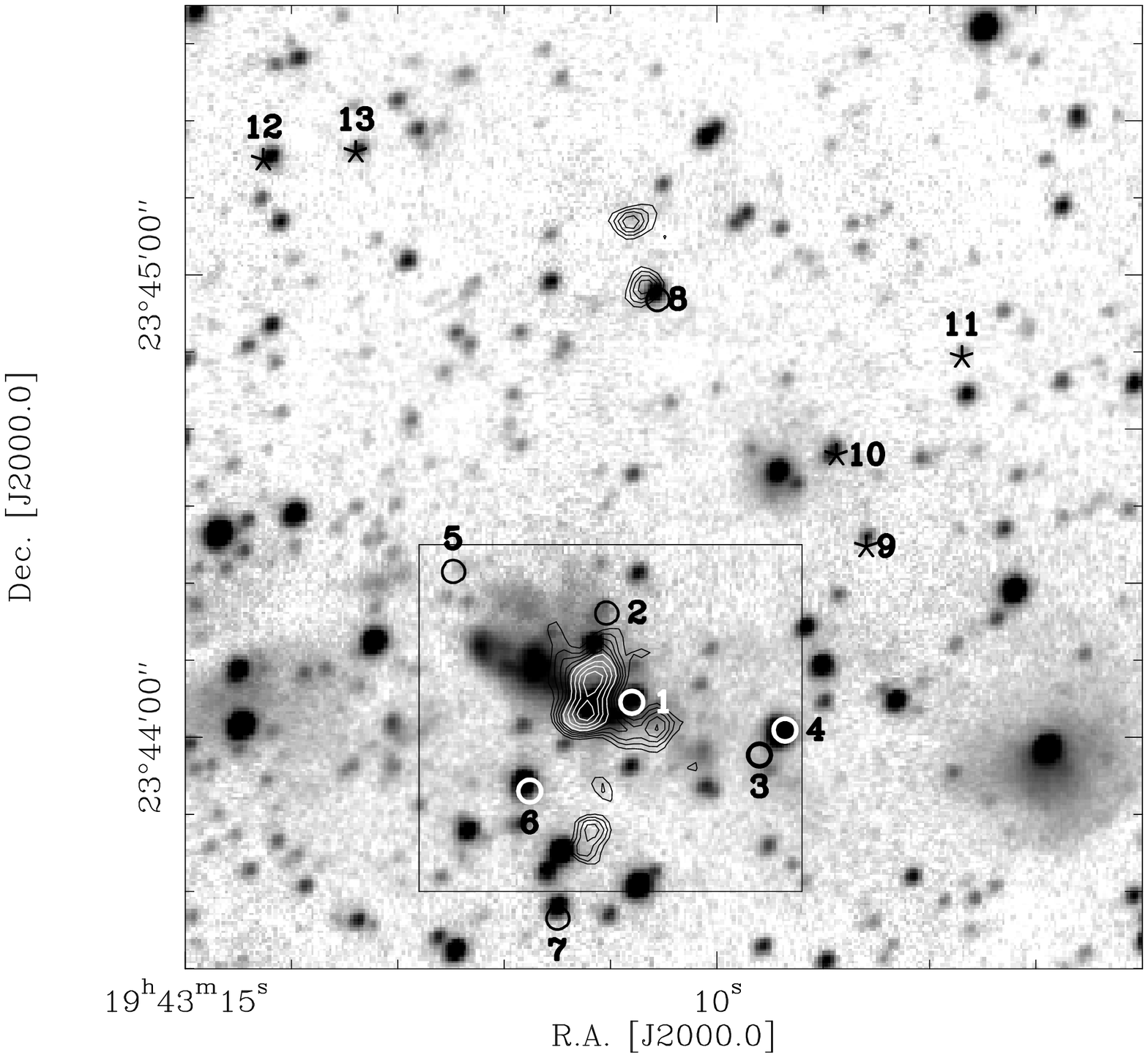}
\caption[K-band, mm and Xray maps]{\footnotesize {\bf top:} The 
contours present the large-scale 1.2~mm dust continuum emission from
$10\%$ to $90\%$ (steps of $10\%$) of the peak flux \citep{beuther
2002a}. The grey-scale shows the K-band image. {\bf bottom:} The
contours show the PdBI 2.6~mm continuum dust cores (black: levels
3.6(1.2=$1\sigma$)8.4~mJy/beam, white: 9.6(2.4)20~mJy/beam) superposed
on the K-band image. Circles and stars mark the X-ray sources
presented in this paper. Sources with circles are most likely
associated with the massive star-forming region, whereas the 
asterisks represent those that might be foreground objects. The box
outlines the region presented in
Fig. \ref{zoomxray}. \label{xrayimage}}
\end{figure}

Here, we present a {\it Chandra} X-ray study of the very young, massive and
deeply embedded star-forming cluster IRAS 19410+2336.  IRAS 19410+2336
is part of a large sample of 69 high-mass protostellar candidates
which has been studied extensively in a series of papers during the
last years
\citep{sridha,beuther 2002a,beuther 2002b,beuther 2002c,beuther 2002d}. 
We assume the source to be located at its near kinematic distance of
2.1~kpc \citep{sridha}, because the derived outflow parameters are
unreasonably high for the far kinematic distance \citep{beuther
2002b}. At the distance of 2.1~kpc, its infrared derived luminosity is
$10^4$~L$_\odot$, and one observes two adjacent star-forming cores
with masses of 840~M$_{\odot}$ and 190~M$_{\odot}$
\citep{sridha,beuther 2002a}. Each core drives a massive bipolar
outflow in east-west direction
\citep{beuther 2002b}. At the center of the southern massive core, a
very compact and weak ($\sim1$\,mJy) cm wavelength source is detected,
which coincides with H$_2$O and Class {\sc ii} CH$_3$OH maser emission
\citep{beuther 2002c}. Figure \ref{xrayimage}(top) gives an overview of 
the region of interest with the 1.2~mm dust continuum data
\citep{beuther 2002a} superposed on an infrared K-band image (\S
\ref{kband}). As the source is located in the Galactic plane, 
confusion due to foreground and background sources is expected and has
to be disentangled by the different observations.  We focus on the
X-ray emission of this region and correlate the detected X-ray sources
with high-resolution images in the mm and near-infrared regime.
\S 2 describes the different observations we performed (X-ray, near-infrared and mm data), and in \S 3 we derive the main physical
parameters of this star-forming region from our data (source
detections, spectra, X-ray luminosities, plasma temperatures and
masses). Finally, \S 4 compiles our conclusions and puts the
observational findings into a more general framework of high-mass star
formation.

%__________________________________________________________________

\section{Observations}

\subsection{{\it Chandra} X-ray observations and data reduction}
\label{obs_chandra}

IRAS 19410+2336 was observed with the ACIS-S3 chip on board the X-ray
telescope {\it Chandra} for 20~ksec on October 15 2001. The ACIS-S3
aim-point was centered at the IRAS position R.A. 19:43:11.4,
Dec. 23:44:06.0. (J2000.0) that is coincident with the main mm
emission peak. The field of interest shown in Fig. \ref{xrayimage} is
much smaller than the $\sim 8' \times 8'$ field of view of the S3
chip, and the FWHM of the point-spread-function of this region is
approximately $0.5''$. The data were reduced with the CIAO 2.2
software package using the CALDB 2.10 database. Both packages are
provided by the {\it Chandra} X-ray
center\footnote{http://asc.harvard.edu/ciao/download.html}. The basic
data product of our observation is the level 2 processed event list
provided by the pipeline processing at the {\it Chandra} X-ray center.

Light curves were extracted for the whole S3 chip as well as just for
the region presented in Figure \ref{xrayimage}. No flares or
enhancements due to low energy protons are observed. The whole
observation time is usable as good time interval. The background
emission is about 0.1 cts pixel$^{-1}$, but as one or two photons
could be detected by background fluctuations anyway, we conservatively
use 6 cts as the detection threshold for an X-ray source in this
field. A wavelet based source detection algorithm was applied using
default parameters \citep{freeman 2002}, and 13 sources between 7 and
19 cts are found in the field of interest (Table
\ref{lumxray}). Based on the 1.2~mm continuum map from the 
30\,m telescope~-- assuming optically thin dust emission~--, we can
estimate an average hydrogen column density of $N_{\rm{H}}=5\times
10^{22}$~cm$^{-2}$ of the molecular cloud (for details on the dust
emission see
\citealt{beuther 2002a}). A Raymond-Smith plasma
model \citep{raymond 1977} with a 3~keV plasma, which is commonly used
to model the X-ray emission of very young and embedded
pre-main-sequence sources \citep{feigelson 1999,preibisch 2002}, 6~cts
correspond to an unabsorbed flux limit of $2 \times
10^{-14}$~erg\,cm$^{-2}$\,s$^{-1}$. Based on the log$N$-log$S$
distribution of \citet{giacconi 2001}, in a field as small as ours
less than 0.1 extragalactic background sources are expected. The
source positions in the field of interest are compared with 2MASS
near-infrared data and should be correct within less than $1''$.

\subsection{Plateau de Bure mm observations}

We observed IRAS 19410+2336 in Summer 2001 with the Plateau de Bure
Interferometer (PdBI) at 2.6~mm in the D (with 4 antennas) and C (with
5 antennas) configuration \citep{guilloteau 1992}. The simultaneously
observed 1~mm data were only used for phase corrections because of the
poor Summer weather conditions. The 3~mm receivers were tuned to
115.27~GHz (USB) (centered at the $^{12}$CO $1\to0$ line) with a
sideband rejection of about 5 dB. At this frequency, the typical SSB
system temperature is 300 to 400 K, and the phase noise was below
30$^{\circ}$. Atmospheric phase correction based on the 1.3~mm total
power was applied. For continuum measurements, we placed two 320~MHz
correlator units in the band to cover the largest possible
bandwidths. In this paper we are focusing on the 2.6~mm continuum
data, the CO line observations will be presented elsewhere. Temporal
fluctuations of the amplitude and phase were calibrated with frequent
observations of the quasars 1923+210 and 2023+336. The amplitude scale
was derived from measurements of MWC349 and CRL618, and we estimate
the final flux density accuracy to be $\sim 15\%$. To cover both cores
a mosaic of 10 fields was observed. The final beam size is $4.1''
\times 3.6''$ (P.A. $-108^{\circ}$). We obtain a $3\sigma$ rms of
$\sim 5$~mJy, which corresponds to a mass sensitivity limit of
approximately 15~M$_{\odot}$ assuming optically thin dust emission at
2.6~mm (see, e.g., \citealt{beuther 2002a} for the deviation of dust
parameters from mm observations).

\subsection{Calar Alto near-infrared observations}
\label{kband}

The near-infrared camera Omega Prime on the 3.5~m telescope on Calar
Alto/Spain was used to obtain K$^\prime$ wide field images of the
region around IRAS~19410+2336 in June 2001. At a pixel scale of
$0.4''$\,pixel$^{-1}$, the 1024$\times$1024 pixel array provides a
field-of-view of $\sim 6.8' \times 6.8'$.  A five position dither
pattern was applied to image the field and to allow for a correction
of array defects that were identified from well illuminated flat-field
frames (for the dark pixels) and a dark frame (for the {\it hot}
pixels).  In each dither position, 15 individual exposures of 2~secs
were stacked into one frame of 30~secs total integration time.  For
each position, the thermal background (sky) was computed by median
combining the four frames resulting from the other dither
positions. This sky frame was subtracted from the respective science
frame, which also removes the bias level. The resulting frame was
divided by a flat-field resulting from dome flats (the difference of a
number of frames of the dome illuminated by a tungsten lamp and a
number of frames taken without illumination). The images of the
various dither positions were then registered and averaged into the
final image. The total integration time in the center of the final
image is 2.5~min, which yields a detection limit ($5\sigma$ peak flux)
for point sources of roughly K$^{\prime}$=17.5.  The seeing-limited
angular resolution in the final image is about $1.2''$. Because
the observation was conducted at high air-masses, its photometry is
rather poor. Thus, we used the Omega Prime data for identification
only, and the photometry is taken from the 2MASS catalog.
 
\section{Results}

\subsection{Source detections and identification}

Fig. \ref{xrayimage}(bottom) shows that in the very young massive
star-forming region IRAS~19410+2336 hard X-ray emission from several
point sources is detected. No evidence for extended or diffuse
emission is found within our detection limits (\S
\ref{obs_chandra}). At mm wavelengths, the two massive cores detected with
the IRAM 30\,m single-dish observations at an angular resolution of $11''$
\citep{beuther 2002a} split up into many sub-sources when observed at
higher spatial resolution (Fig. \ref{xrayimage}, bottom), confirming
that a cluster of stars is forming. The interferometric data highlight
the real massive cores, whereas the large-scale surrounding core
emission is filtered out by the interferometric observing technique. A
detailed analysis of the protocluster properties will be presented
elsewhere. Here we are focusing on the spatial associations of X-ray,
near-infrared and mm sources.

\begin{figure}[ht]
\begin{center}
\includegraphics[angle=-90,width=8.7cm]{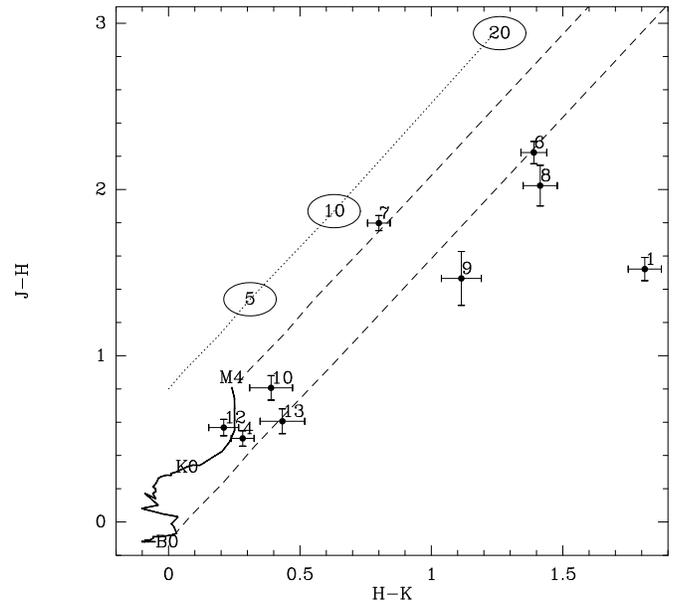}
\end{center}
\caption[Near-infrared color-color diagram]{\footnotesize Near-infrared 
color-color diagram of X-ray associated sources based on 2MASS
data. The numbers correspond to the sources numbered also in
Fig. \ref{xrayimage}. The full line presents main sequence colors with
three labeled spectral star types \citep{ducati 2001}, and the dashed
lines show the reddening band for main sequence colors \citep{rieke
1985}. The dotted line gives the reddening scale with encircled visual
extinctions ([mag]). Source X1 shows the strongest near-infrared
excess.\label{color}}
\end{figure}

Thirteen X-ray sources are detected within the field of interest
(Table \ref{lumxray}). Remarkably, nearly all X-ray sources are,
within the positional uncertainty, associated with near-infrared
counterparts observed in the K-band (X3 and X5 only
tentatively). Because the K-band extinction is approximately only
$10\%$ of the visual extinction, in the K-band we can observe more
embedded regions of star-forming cluster than possible in the optical.
We recall that the extinction in the soft X-ray regime is similar
to the K-band extinction, and that it drops even further going to
higher X-ray energies \citep{casanova 1995,ryter 1996}.  In contrast
to the K-band counterparts, only 2 sources (X2 and X8) are found in
the direct vicinity of mm dust cores that are tracing the coldest and
deepest embedded regions where presumely the youngest sources are
found. The whole region is also covered by the 2 Micron All Sky Survey
(2MASS\footnote{For details see the 2MASS web-site at
http://www.ipac.caltech.edu/2mass/releases/docs.html}) in the
near-infrared bands J, H and K. Except for sources X2, X3, X5 and X11,
which are detected by our K-band Omega Prime observation but
which are too faint to be detected by the 2MASS survey, all other
X-ray sources are also 2MASS point sources. A color-color diagram
helps to determine the characteristics of the associated near-infrared
counterparts (Fig. \ref{color}). While a few of the near-infrared
sources are located around the unreddened main sequence in the
color-color diagram, many sources are reddened and show near-infrared
excess, suggesting that they are pre-main-sequence objects surrounded
by circumstellar material. This is expected in view of the early
evolutionary stage, the large core masses and the high column
densities of the region \citep{beuther 2002a}. Table
\ref{lumxray} compiles reddening estimates of the X-ray near-infrared
counterparts. The $A_{\rm{v}}$ of X1 is most uncertain because X1 has
the largest near-infrared excess which makes the $A_{\rm{v}}$
determination more difficult. Using the reddening scale
(Fig. \ref{color}), we can still estimate an approximate value of the
visual extinction $A_{\rm{v}}$ of X1.

\subsection{Spectral information of the X-ray data}
\label{specinfo}

\subsubsection{X-ray spectra and X-ray luminosities} 
\label{specinfo1}

\begin{figure*}[ht]
\includegraphics[angle=-90,width=18cm]{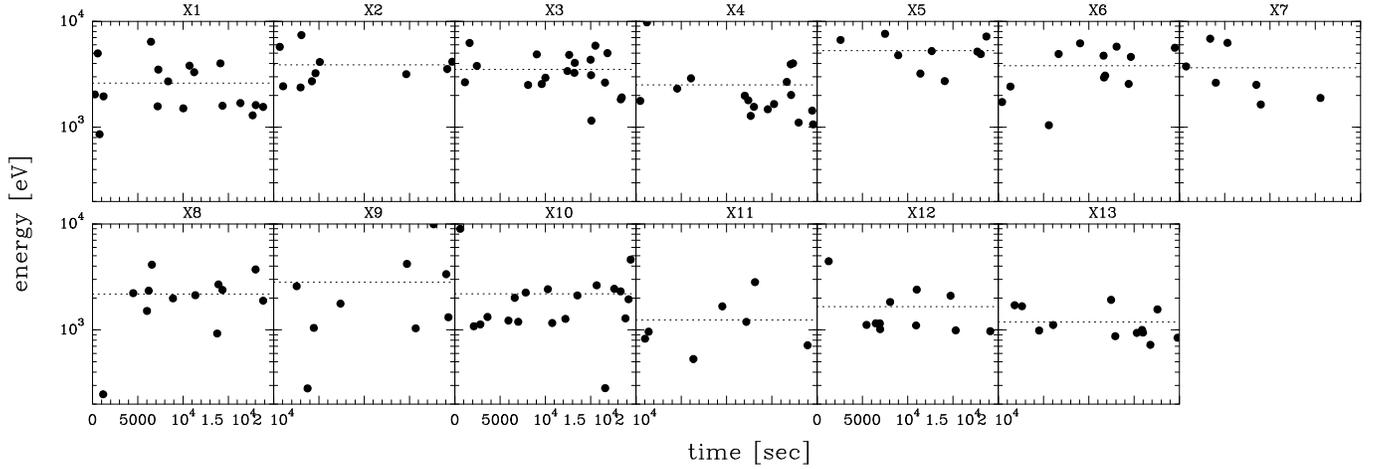}
\caption[Photon energy versus time]{\footnotesize Photon energy versus 
time for the 13 detected sources within the field of interest. The
dotted lines mark the mean detected photon energy in each source (see
also Table
\ref{lumxray}). \label{spectra}}
\end{figure*}

The number of photons per detected source ranges from 7 to 19 counts
within the 20~ksec observation time (Table \ref{lumxray}). We note
that we are dealing with low-number statistics, and this has to be
taken into account when deriving the physical parameters of this
region. Figure \ref{spectra} presents the photon energies of the 13
X-ray sources versus time.  As expected, we detect emission mainly
between 2~keV and 6~keV in the near-infrared reddened and thus
embedded sources, whereas the sources, that are near main sequence
colors in the color-color diagram, emit more in the soft regime below
2~keV (see also mean values in Figure \ref{spectra} and Table
\ref{lumxray}). The deficiency of soft photons ($<2$~keV) in the
direction of the embedded sources is most probably due to the high
column densities that absorb the soft X-rays. The lack of photons
above 6~keV may be a real feature of the X-ray sources, but it has
also to be noted that the sensitivity of {\it Chandra} above 6~keV decreases
significantly. The effective collection area of {\it Chandra} at 8~keV drops
to less than a quarter of that at 4~keV\footnote{See Chandra
Proposers' Observatory Guide, 2000, http://cxc.harvard.edu}. The fact
that we detect hard X-ray emission ($>3$~keV) of the embedded sources
fits in the general finding that the younger the sources are, the
harder the X-ray emission is (e.g., \citealt{feigelson 1999} or \S
\ref{introxray}). The X-ray spectra of X1 to X7 show all such hard
X-ray signatures, which makes them likely to belong to the forming
cluster. The distribution of photon energies cannot be explained by a
low-temperature-Raymond-Smith plasma of 1-2~keV \citep{raymond 1977},
usually used to fit the X-ray spectra of low-mass T Tauri stars
\citep{feigelson 1999}, because the column densities $N_{\rm{H}}$,
that are necessary to attenuate all the emission below
2~keV($N_{\rm{H}}$ a few times $10^{23}$~cm$^{-2}$), would imply
average X-ray luminosities as high as $10^{34}$~erg\,s$^{-1}$. This
value exceeds the known luminosities of young stellar objects by
orders of magnitude (\S \ref{introxray}), and it could hardly be
reached even during strongest flaring events (e.g.,
\citealt{preibisch 1993}, \citealt{grosso 1997}, \citealt{tsuboi 2000}). 
Contrasting these hard spectra, X8 to X13 show softer spectra. While
some of them might be unrelated objects, others could belong to the
cluster but be in a more advanced state of evolution (for details see
\S \ref{detail}).

In order to derive estimates of the X-ray plasma temperatures and the
X-ray luminosities, we performed a detailed spectral analysis.  We
extracted individual pulse-height spectra and built the
corresponding redistribution matrices and ancillary response files for
each source.  We fitted the ungrouped pulse-height spectra with XSPEC,
using a single-temperature thermal plasma model (model ``raymond'',
\citealt{raymond 1977}) plus the absorption model ``wabs''.  The
plasma temperatures and column densities derived in these fits are
compiled in Table \ref{lumxray}. The fitting results were then used to
integrate (in XSPEC) the model fluxes over the 0.2--10 keV band and to
compute the X-ray luminosities.  In some cases the low number of
detected photons allowed only to derive lower limits to the plasma
temperature.  We are fully aware that, given the rather low number of
photons per source, the derived spectral parameters inherit a
relatively large statistical uncertainty. Nevertheless, we believe
that the fitting results contain valuable information, for the
following three reasons:

(1) The spectral parameters derived in the fits allow a clear
distinction of two different groups among the sources: sources X1-X7
show high plasma temperatures (k$T>3$\,keV) and high colum densities
($N_{\rm{H}}>10^{22}$\,cm$^{-2}$); in contrast, the sources X8-X13
show low plasma temperatures (k$T<1.5$\,keV) and (with the exception
of X8) low colum densities ($N_{\rm{H}}<10^{22}$\,cm$^{-2}$).  These
two groups exactly reflect the spatial distribution of the sources:
X1-X7 are located inside the central region of the molecular clump,
while X8-X13 lie outside the molecular clump (Fig. \ref{xrayimage}).

(2) The column densities derived in the
spectral fits agree in nearly all cases rather well with expectations
from the infrared colors of the sources (Table
\ref{lumxray}, $A_{\rm{v}}=N_{\rm{H}}/2\times10^{21}$).

(3) \citet{getman 2002} have performed fitting simulations of
pulse-height spectra with low numbers of counts to estimate the
statistical uncertainties of the fitting results.  They found that
even if the column densities and temperatures derived in the fits are
subject to rather large uncertainties, the broad-band X-ray
luminosities derived from the fitting results are quite reliable
(statistical error $\le 35\%$ for spectra with 15 counts).

Very likely, the X-ray emission of these objects originates from a hot
thermal plasma typical for very young embedded objects
(e.g., \citealt{feigelson 1999}, \citealt{preibisch 2002},
\citealt{kohno 2002}). But as the number of photons and by that the 
spectroscopic information is small, other X-ray emission processes
have to be taken into account, e.g., non-thermal power-law
distributions as a signature of synchrotron emission, which has been
observed in the radio regime by \citet{reid 1995} in the massive
star-forming region W3(OH). Therefore, we also estimate X-ray
luminosities with three different power-law distributions
$S_{\rm{x}}\propto E^{-\alpha}$ ($\alpha=0.5,1,2$) to derive an estimate
for the influence of $\alpha$ on the total X-ray luminosity (using the
web-based PIMMS
tool\footnote{http://asc.harvard.edu/toolkit/pimms.jsp}). The derived
luminosities vary by less than a factor 1.5 for the different $\alpha$.
As hot plasmas are dominated by bremsstrahlung above k$T\sim 1$~keV
\citep{fink 1982}, which can also be approximated by non-thermal
power-law distributions, the X-ray luminosities derived this way are
close to the luminosities calculated using the Raymond-Smith plasma
(they differ only a factor 2 at maximum).

We calculated $L_{\rm{x}}$ only for X1 to X8 (Table \ref{lumxray}),
because the other sources could be foreground or background objects,
and thus no reliable distance is known. We find X-ray luminosities in
the $10^{31}$~erg\,s$^{-1}$ range. These values are high, and
above or at the upper end of classical T Tauri X-ray luminosities,
i.e., in the regime of younger Class I sources \citep{feigelson 1999}
and/or more massive Herbig Ae/Be type objects \citep{zinnecker
1994,gregorio 1998,hamaguchi 2000}. Unresolved binary systems could
alter the results slightly, but not by orders of magnitude.

\begin{table*}[ht]
\caption[X-ray luminosities]{\footnotesize Compilation of the 
parameters derived from the X-ray and 2MASS data: positions of the
X-ray sources, number of counts per source, mean energy of detected
photons, best fits of Raymond-Smith plasma temperatures (k$T$\,[keV]
and $T$\,[K]) and derived X-ray luminosities $L_{\rm{X}}$ with error
ranges. Colons mark less conclusive fits.  Additionally, the column
densities $N_{\rm{H}}$ (and error ranges) from the X-ray fits and the
2MASS derived extinctions $A_{\rm{v}}$ are presented.}
\label{lumxray}
\begin{center}
\begin{tabular}{lrrrrrrrrr}
%\hline
source & R.A. & Dec. & \# & mean & k$T$ & $T$ & $L_{\rm{X}}$ & $N_{\rm{H}}$& $A_{\rm{v}}$\\ 
& (J2000.0) & (J2000.0) & [cts] & [keV] & [keV] & [$10^7$\,K] &
[$10^{31}$erg\,s$^{-1}$] & [$10^{22}$cm$^{-2}$] & \\
\hline
X1 & 19:43:10.8 & +23:44:04.6  & 17 & 2.6  & 5.6  [$>4$ ]     & 6.5  [$>4.6$ ] & 1.1 [0.7--1.4]& 1.0 [0.7--1.3] & 17: \\
X2 & 19:43:11.0 & +23:44:16.1  & 10 & 3.9  & 9.6  [$>6$ ]     & 11.1 [$>7$ ]   & 2.1 [1.5--2.8]& 5.9 [4.5--8.0] & --  \\
X3 & 19:43:09.6 & +23:43:57.7  & 19 & 3.5  & 10:  [$>6$ ]     & 11.6:[$>7$ ]   & 2.4 [1.9--3.0]& 2.9 [2.3--3.8] & --  \\
X4 & 19:43:09.4 & +23:44:01.0  & 17 & 2.5  & 3.2  [2.2--5.3]  & 3.7  [2.6--6.2]& 1.2 [0.8--1.7]& 1.1 [0.9--1.4] & --  \\
X5 & 19:43:12.5 & +23:44:21.8  &  9 & 5.3  & 9.3  [$>5$ ]     & 10.8 [$>5.8$ ] & 2.6 [1.6--3.5]& 10: [$>7.2$ ]  & --  \\
X6 & 19:43:11.8 & +23:43:53.0  & 12 & 3.8  & 7.4  [$>5$ ]     & 8.6  [$>5.8$ ] & 1.8 [1.2--2.3]& 3.7 [2.6--5.3] & 18  \\
X7 & 19:43:11.5 & +23:43:36.5  &  7 & 3.6  & 7.5  [$>4$ ]     & 8.7  [$>4.6$ ] & 0.8 [0.5--1.1]& 2.7 [1.8--3.9] & 12  \\
X8 & 19:43:10.6 & +23:44:56.8  & 12 & 2.2  & 1.4  [1.1--1.9]  & 1.6  [1.3--2.2]& 2.2 [2.0--2.8]& 3.0 [2.5--3.7] & 17  \\
X9 & 19:43:08.6 & +23:44:24.7  &  9 & 2.8  & 1.2:             & 1.4:           & --            & 0.3:           & 12  \\
X10& 19:43:08.9 & +23:44:36.6  & 19 & 2.2  & 1.5:             & 1.7:           & --            & 0.8:           &  3  \\
X11& 19:43:07.7 & +23:44:49.3  &  7 & 1.2  & 1.0:             & 1.2:           & --            & 0.2:           & -- \\
X12& 19:43:14.3 & +23:45:14.9  & 12 & 1.7  & 1.3  [1.1--1.7]  & 1.5  [1.3--2.0]& --            & 0.6 [0.4--0.8] &  1  \\
X13& 19:43:13.4 & +23:45:15.9  & 13 & 1.2  & 1.3  [1.1--1.5]  & 1.5  [1.3--1.7]& --            & 0.3 [0.2--0.4] &  2  \\
\hline
\end{tabular}
\end{center}
\end{table*}

% Ergebnisse der neuen Fits vom 10.7.2002
%    &   kT           &  N_{\rm H}        &  L_{\rm X}
%                        /1.e22               /1.e31
%X1    5.6  [>4 ]        1.0 [0.7--1.3]          1.1 [0.7--1.4]
%X2    9.6  [>6 ]        5.9 [4.5--8.0]          2.1 [1.5--2.8]
%X3    10:  [>6 ]        2.9 [2.3--3.8]          2.4 [1.9--3.0]
%X4    3.2  [2.2--5.3]   1.1 [0.9--1.4]          1.2 [0.8--1.7]
%X5    9.3  [>5 ]        10: [>7.2 ]             2.6 [1.6--3.5]
%X6    7.4  [>5 ]        3.7 [2.6--5.3]          1.8 [1.2--2.3]
%X7    7.5  [>4 ]        2.7 [1.8--3.9]          0.8 [0.5--1.1]
%X8    1.4  [1.1--1.9]   3.0 [2.5--3.7]          2.2 [2.0--2.8]
%X9    1.2:              0.3:
%X10   1.5:              0.8:
%X11   1.0:              0.2:
%X12   1.3  [1.1--1.7]   0.6 [0.4--0.8]
%X13   1.3  [1.1--1.5]   0.3 [0.2--0.4]

\subsubsection{Variability}

We do not see strong intensity flaring in our observations
(Fig. \ref{spectra}), but sources X2, X3 and X4 show absence of
detected photons in the time interval between $6\times 10^3$ and
$10^4$ sec, which indicates possible variability in these sources. The
other sources show a rather constant count rate within the 20~ksec of
our observation. To quantify the possible variability, we performed a
statistical test of the photon arrival times following the approach
outlined by \citet{preibisch 2002}: given the number of detected
photons in one source in the total integration time, we compute the
mean count rate and specify the time period in which two photons are
expected assuming a constant count rate. Then we determine the maximum
number of photons $N_{\rm{max}}$ detected in this source within any
time interval as long as the one derived for the two photons. The
Poisson probability to find $N_{\rm{max}}$ or more counts due to
statistical fluctuations in a period for which two photons are
expected, is given by

$$P=1-\sum_{k=0}^{N_{\rm{max}}-1}~e^{-2} \frac{2^k}{k!}.$$

The {\it probability of variability} ($pov$) is given by $pov=1-P$ (for
details see, e.g., \citealt{bevington 1992}). The $pov$ values for
X3 and X4 are only $68\%$ and $85\%$, giving
just marginal evidence for variability within our total observing
time. Contrasting to that, $pov$ equals $98.3\%$ for X2,
therefore this source is consistent with X-ray variability.

\subsection{Mass estimates from the near-infrared data}
\label{mass_estimates}

The 2MASS near-infrared K-band data provide a tool to estimate the
masses of the infrared sources. Taking the observed K-band
brightnesses $m_{\rm{K}}$ at the given K-band extinction
($A_{\rm{K}}\sim 0.11\times A_{\rm{v}}$) we get an estimate of the
intrinsic brightness $M_{\rm{K}}$ of the source assuming the distance
$d=2$~kpc:

$$m_{\rm{K}}-M_{\rm{K}}=5\, {\rm{log}} (d)-5+A_{\rm{K}}.$$

Different pre-main-sequence tracks are discussed in the
literature (e.g., \citealt{dantona 1994}, \citealt{baraffe 1998},
\citealt{palla 1999}, \citealt{siess 2000}, \citealt{dantona 2000}).
While some of them differ significantly at the low-mass end, they are
better comparable going to masses $>1$\,M$_{\odot}$ \citep{siess
2000}. Most of the calculated tracks do not cover stars more massive
than 2.5\,M$_{\odot}$, but \citet{palla 1999} calculate the
pre-main-sequence evolution to 6\,M$_{\odot}$ and \citet{siess 2000}
to 7\,M$_{\odot}$. At the high-mass end, which we are particularly
interested in, both calculations agree well. The tracks of
intermediate-mass stars ($>2$\,M$_{\odot}$) are following almost 
horizontal tracks of equal luminosity in the Hertzsprung-Russel
diagram, and the luminosity is almost independent of the
temperature. Thus, it is possible to estimate luminosities and by
that masses from our K-band observations, but the data do not give any
information about the age of the objects. \citet{siess 2000} stress
that any age determination below $10^6$\,yr is highly uncertain. In
the following, we use the tracks compiled by
\citet{palla 1999}, but the results are similar when using the
tracks of \citet{siess 2000}.

Comparing the intrinsic K-band brightnesses $M_{\rm{K}}$ with the
theoretical tracks by \citet{palla 1999}, we can derive mass estimates
for the near-infrared objects. We apply this procedure for the
near-infrared sources corresponding to X1, X6, X7 and X8, because only
those are 2MASS detections and likely belong to the forming massive
clusters (\S \ref{specinfo}). The pre-main-sequence tracks provided by
\citet{palla 1999} for stars up to 6~M$_{\odot}$ do not comprise
sources intrinsically brighter than $-1.6$~mag in the K-band. Sources
more massive than 8~M$_{\odot}$ do not have an optically visible
pre-main-sequence phase, because they start nuclear burning already
being deeply embedded in their natal cores \citep{palla
1993}. Therefore, X6 and X7 could be more massive than 6~M$_{\odot}$,
but it has to be noted that the derived model brightnesses
$M_{\rm{K}}$ neglect luminosity due to the disk, i.e., heated dust in
the disk, and accretion luminosity within the disk and between the
disk and the protostar. \citet{palla 1999} stress that the
understanding of these processes is still inadequate for quantitative
predictions, but as the near-infrared excess in X6, X7 and X8 is not
very high (Fig. \ref{color}), to our judgment the additional disk
luminosity in these sources is not the dominant effect. This is
different for source X1 which has considerable infrared excess and
makes K-band brightnesses difficult to be compared with theoretical
predictions. But as stars below 2~M$_{\odot}$ are at least 2
magnitudes fainter in the theoretical predictions than the observed
brightness of X1, 2~M$_{\odot}$ can still be regarded as a tentative
lower mass limit for this source. Real upper mass limits are given by
the total bolometric luminosity of the whole cluster of $\sim
10^4$~L$_{\odot}$
\citep{sridha}. A luminosity of $10^4$~L$_{\odot}$ corresponds in 
the case of main sequence stars to a B1 star (13\,M$_{\odot}$). As a
cluster of stars is forming in IRAS~19410+2336, the mass distribution
of the cluster members is likely to follow an initial mass
function. In spite of their larger number, the luminosity-mass
relation for low-mass stars is lower ($L\propto m^{2.8}$ for
$m<1$\,M$_{\odot}$) than for high-mass stars ($L\propto m^{4}$ for
$1<m<30$\,M$_{\odot}$, \citealt{schatzman 1993}), and still the major
contribution of the total luminosity stems from the massive cluster
members. For a detailed IMF discussion of the initial source sample
see \citet{sridha}. As the most massive object is likely the central
mm source, which is not detected in X-ray emission, the X-ray detected
sources are $\leq 10$~M$_{\odot}$. Additionally, the models give
the approximate luminosity of the sources and we get an estimate of
the $L_{\rm{X}}/L_{\rm{bol}}$ ratio ($L_{\rm{X}}$ is taken from \S
\ref{specinfo} and Table \ref{lumxray}). The parameters and results
are listed in Table
\ref{ir_masses}.

\begin{table}[ht]
\caption[Near-infrared derived masses]{\footnotesize Near-infrared parameters of most likely cluster sources with 2MASS counterparts. Given are masses ($\leq 10$~M$_{\odot}$), bolometric luminosities $L_{\rm{bol}}$, and the ratio  $L_{\rm{X}}/L_{\rm{bol}}$ \label{ir_masses}}
\begin{center}
\begin{tabular}{lrrrrrrr}
%\hline
src & $m_{\rm{K}}$ & $A_{\rm{v}}$ & $A_{\rm{K}}$ & $M_{\rm{K}}$ & mass
& $L_{\rm{bol}}$ & $L_{\rm{X}}/L_{\rm{bol}}$ \\ & [mag] & & & [mag]
&[M$_{\odot}$] & [L$_{\odot}$] & [$10^{-5}$]\\
\hline
X1  & 11.7 & 17 & 1.7  & $-1.5$  & $>2$ &   $>10$        & $<70$ \\
X6  & 11.0 & 18 & 1.8  & $-2.3$  & $>6$     & --         & --    \\
X7  & 8.8  & 12 & 1.2  & $-3.9$  & $>6$     & --         & --    \\
X8  & 12.8 & 17 & 1.7  & $-0.4$  & $>3$   & $>30$        & $<2$  \\
\hline
\end{tabular}
\end{center}
\end{table}

The derived masses are in the intermediate-mass regime of Herbig Ae/Be
stars, clearly more massive than usual T Tauri or Class I sources. 
\citet{zinnecker 1994} observed X-ray emission from Herbig
Ae/Be stars, and those sources showed rather soft
spectra. \citet{preibisch 1995} speculate that the observations could
be explained by coronal emission due to a shear dynamo. As the
spectral signatures of the sources we are studying here are
significantly harder than those found in their sample (\S
\ref{specinfo}), we suggest that the X-ray emission could be
due to magnetic star--disk interactions as proposed for very young
Class I sources \citep{montmerle 2000}. In that framework, it is
possible that the sources observed in IRAS~19410+2336 are precursors
of the better studied Herbig Ae/Be stars. The possible
$L_{\rm{X}}/L_{\rm{bol}}$ ratio-regime (Table \ref{ir_masses}) for two
sources is at the upper boundary consistent with ratios found in very
young low-mass objects ($\sim 2\times 10^{-4}$, e.g.,
\citealt{feigelson 1999}, \citealt{preibisch 2002}), but at the lower 
end it is also consistent with the $L_{\rm{X}}/L_{\rm{bol}}$ ratios
found in samples of Herbig Ae/Be stars \citep{zinnecker 1994,gregorio
1998}. It could be argued that the X-ray emission is due to low-mass
counterparts, but as the derived X-ray luminosities are significantly
larger than typical values in the low-mass regime, most of the X-ray
emission stems likely from the intermediate-mass sources.

\subsection{The X-ray sources in detail}
\label{detail}

\paragraph{Source X1 and the main mm core:} The most 
remarkable X-ray source is X1, which does not only show strong
reddening but also the largest infrared excess indicative embedded
protostellar objects \citep{feigelson 1999}. As the infrared derived
mass is $\geq 2$~M$_{\odot}$ but the X-ray spectrum harder than usual
Herbig Ae/Be stars \citep{zinnecker 1994}, X1 is probably not a normal
Herbig Ae/Be star but a precursor of such an object.

\begin{figure}[ht]
\begin{center}
\includegraphics[angle=-90,width=8.7cm]{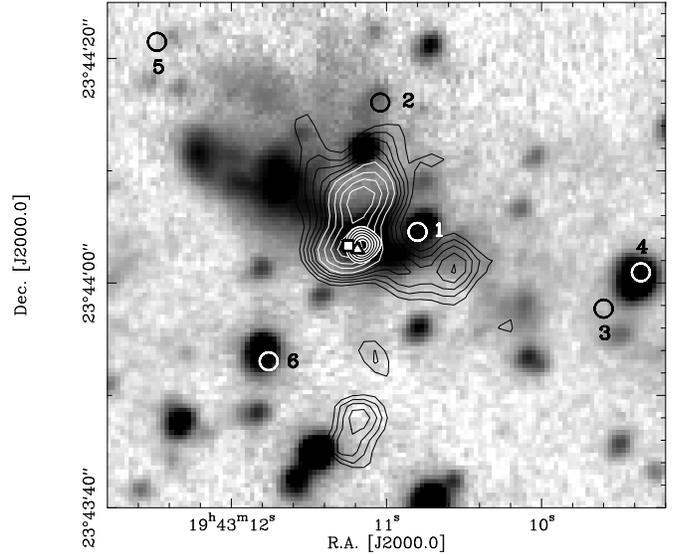}
\end{center}
\caption[Zoom into the center]{\footnotesize Zoom into the center 
of the field presented in Fig. \ref{xrayimage}. Contours and symbols
are the same as in Fig. \ref{xrayimage}. Additionally, the black
contours on white ground right at the center show the cm-peak, the
triangle pinpoints the H$_2$O maser position and the square the
CH$_3$OH maser positions
\citep{minier 2001,beuther 2002c}. \label{zoomxray}}
\end{figure}

Source X1 is the X-ray source nearest to the core center but not
coincident with a mm core. Figure \ref{zoomxray} shows that the main
mm core coincides with a compact cm source as well as with H$_2$O and
Class {\sc ii} CH$_3$OH maser emission \citep{minier 2001,beuther
2002c}, while the X-ray source is clearly offset and coincides with a
near-infrared source $\sim 6''$ to the west. Thus, the main mm core is
likely to contain the youngest and most massive object of the cluster,
and we do not detect X-ray emission from this source, but only from
another object nearby that might be slightly more evolved. Therefore,
we do detect X-ray emission from pre-main-sequence sources but not
from the youngest and massive center of the cluster. Assuming
optically thin dust emission, we estimate the core mass of the central
source from the 2.6~mm continuum data to about 80~M$_{\odot}$ and the
H$_2$ column density to $\approx 2\times 10^{24}$~cm$^{-2}$,
corresponding to a visual extinction $A_{\rm{v}}$ of approximately
2000 (for details on mm dust calculations see 
\citealt{beuther 2002a}). Taking into account this visual extinction 
and assuming different Raymond-Smith plasma temperatures for the X-ray
emission, we can calculate a range of upper limits for the X-ray
luminosity of this source: a 3~keV plasma temperature corresponds to
an upper limit of $L_{\rm{X}}<9\times 10^{34}$~erg\,s$^{-1}$, while a
temperature of 10~keV implies $L_{\rm{X}}<4\times
10^{34}$~erg\,s$^{-1}$. A lower plasma temperature of 1~keV results in
an upper limit of $L_{\rm{X}}<14\times 10^{36}$~erg\,s$^{-1}$.
However, it has to be noted that the possibility remains that there is
no significant X-ray emission in the earliest stages of massive star
formation.

\paragraph{Sources X2 and X5:}
Source X2 is at the edge of the southern mm emission and associated
with some faint K-band emission that might be a reflection nebula from
the main sources to the south. As the K-band emission is weak, it is
not detected by 2MASS. Being in the near vicinity of a mm core and
showing a very hard X-ray spectrum, it is possible that X2 might be
powered by an embedded protostellar object. It has to be noted that
this source is the only one with observed X-ray variability. Source X5
is also too faint in the near-infrared to be detected by 2MASS, and it
is difficult to decide whether the nearby faint K-band emission is a
stellar feature or a reflection nebulae. But as X5 has
the hardest X-ray spectrum of our sample, it is very likely a young
protostellar object within the cluster of IRAS~19410+2336.

\paragraph{Sources X3 and X4:}
These sources are both close to the same near-infrared source, but
accurate astrometry indicates that X4 rather than X3 is associated
with it (Fig. \ref{zoomxray}). Additionally, we find an optical
counterpart for X4 in the Digitized Sky Survey provided by the Space
Telescope Science Institute. Based on the X-ray spectrum, which is not
as hard as the spectra of the other cluster members, and the optical
and infrared data, we cannot distinguish whether X4 belongs to the
cluster or is an unrelated foreground object. In contrast, X3 might be
associated with more diffuse K-band emission (Fig. \ref{zoomxray}). As
the hard X-ray spectrum of X3 is indicating an early evolutionary
stage, this source most likely is a very young pre-main-sequence
object belonging to the cluster under investigation.

\paragraph{Sources X6 and X7:}
These sources also show hard X-ray spectra, and both are within the
large-scale region of the southern massive core as seen in
Fig. \ref{xrayimage}(top), but no mm core
corresponds to any of them. This is reflected in the color-color
diagram, where they inhabit a reddened region without strong
infrared excess. The infrared-derived mass estimates are not
conclusive, but the data~-- X-ray and near-infrared~-- are consistent
with precursors of Herbig Ae/Be stars of masses between 5~M$_{\odot}$
and 10~M$_{\odot}$.

\paragraph{Source X8:}
Source X8 is associated with the strongest mm peak of the northern
cluster as well as with a near-infrared and even an optical
counterpart (as found in the Digitized Sky Survey). The 2MASS data
show that the infrared source is reddened and has infrared excess, and
we derive a lower mass limit of approximately 3.5~M$_{\odot}$. The
X-ray spectrum is softer than for the southern cluster members, thus
it is possible that X8 is in a slightly more evolved state of
evolution. 

\paragraph{Sources X9 and X10:}
The X-ray spectrum of these sources is softer, but they are still
reddened in the color-color diagram and X9 even shows infrared
excess. It is likely that they are also young stellar objects, but we
are not able to determine whether they belong to the cluster or
whether they are foreground objects.

\paragraph{Sources X11, X12 and X13:}
These sources also have a rather soft spectrum, and the color-color
diagram locates them on or near the unreddened main sequence
(X11 is too weak to be a 2MASS detection). As they are
also spatially offset from the large-scale mm core, we regard them as
not associated with the massive star-forming region.

\section{Conclusions}
 
Hard X-ray emission from a number of point sources is detected in a
young, massive and embedded star-forming cluster in a very early stage
of evolution. We did not detect X-ray emission from the most massive
and central object (upper limit $L_{\rm{X}}<9 \times
10^{34}$~erg\,s$^{-1}$ for k$T=3\,$keV) but from a few
sources in its vicinity. Typical X-ray properties of high-mass
main-sequence stars ($M_\star \geq 8\,M_\odot$), where the emission
originates from internal shocks in the radiation-driven stellar winds,
are $L_{\rm X}/L_{\rm bol} \sim 10^{-7}$, soft X-ray spectra with
typical temperatures of k$T \la 0.5$~keV and very little variability
\citep{berghoefer 1997}. The X-ray properties we find for X1 to X7 are
clearly different, they show hard X-ray emission (k$T \geq 3$~keV)
and most of them also near-infrared excess, very similar to those
observed for extremely young embedded objects (Class I protostars),
which have typical plasma temperatures of k$T \sim 5-10$~keV
\citep{feigelson 1999,imanishi 2001}.  Combining infrared data with
pre-main-sequence evolutionary tracks
\citep{palla 1999}, it is possible to estimate the approximate masses
of some of the hard X-ray sources. Those estimates indicate that they
are in the intermediate-mass regime of Herbig Ae/Be objects. Taking
additionally into account the harder X-ray spectra compared with other
Herbig Ae/Be studies, it is likely that the X-ray sources in
IRAS~19410+2336 are even precursors of Herbig Ae/Be stars. The
emitted X-ray photons with energies mostly above 2~keV indicate plasma
temperatures $>10^7$~K and X-ray luminosities around a few times
$10^{31}$~erg\,s$^{-1}$. The latter values are well within the regime of Class
I low-mass protostars \citep{feigelson 1999}, but they are also
consistent with the results obtained for Herbig Ae/Be stars
\citep{zinnecker 1994}. Thus, some of the objects are probably very
young intermediate-mass pre-main-sequence sources, whereas other
sources could also be low-mass Class I or T Tauri stars. The emission of
one of the sources is consistent with X-ray variability.

In spite of the observation of hard X-ray emission in the
weak-lined T Tauri star V773 \citep{tsuboi 1998}, where the disk has
already been dissipated to a large degree, it is unlikely that the
hard X-ray spectra observed in younger class I sources are due to
enhanced solar-type magnetic activity. Therefore, it is proposed that
the hard X-ray emission, which is more often observed in class I
sources than in weak-lined T Tauri stars, is produced by magnetic
reconnection effects between the protostars and their accretion disks
\citep{hayashi 1996,feigelson 1999,montmerle 2000}. As the X-ray
spectra of the intermediate-mass objects in IRAS~19410+2336 exhibit
very similar signatures to such low-mass sources, our results are
consistent with disks being present in intermediate-mass star
formation as well.

For a better understanding of the nature of the underlying X-ray
powering sources much work has to be done in the future. Deeper X-ray
and near-infrared images will help to set stronger constraints on the
physical properties of the sources: it will be necessary to obtain
sensitive X-ray spectra to determine better the absorbing $N_{\rm{H}}$
column densities and plasma parameters. It is also of great interest
to further investigate the properties of the central and deepest
embedded object, which means lowering the detection
limits. Furthermore, the variability of the X-ray sources in very
young massive star-forming regions is not known so far. Therefore,
several approaches should be followed in the years coming: deep
{\it Chandra} observations of the source of interest will disclose
variabilities and faint emission of the central object. Additionally,
a sample of similar sources has to be identified, because only a
statistical analysis of several young high-mass star-forming regions
can build a solid picture of the relevant physical processes. As high
spatial resolution is essential for many of this studies, {\it Chandra} is a
very promising choice. But considering the higher sensitivity of
XMM-Newton, it might be possible to study grating X-ray spectra of the
brightest sources of the sample of clusters studied then. On the
near-infrared side, we suggest to get deeper images in the J, H and K
bands to improve the mass estimates of the X-ray emitting sources, and
near-infrared spectroscopy might help classifying the types of stars
\citep{hanson 2002}. 

To summarize, X-ray studies of young massive star-forming regions are
just in its infancy, and the next years with the space telescopes
{\it Chandra} and XMM-Newton will bring many new insights in that research
area. We also like to stress that multi-frequency studies over a wide
range of bands are extremely promising approaches for the
understanding of the physical processes forming massive stars.

\begin{acknowledgements}

We thank an anonymous referee for very helpful and detailed
comments on the paper. This publication makes use of data
products from the Two Micron All Sky Survey, which is a joint project
of the University of Massachusetts and the Infrared Processing and
Analysis Center, funded by the National Aeronautics and Space
Administration and the National Science Foundation. We also used data
from the Digitized Sky Survey as provided by the Space Telescope
Science Institute.

\end{acknowledgements}

\end{document}